\documentclass{article}
\usepackage[numbers,sort&compress]{natbib}

\usepackage[final]{neurips_2024}

\usepackage[utf8]{inputenc} 
\usepackage[T1]{fontenc}    
\usepackage{hyperref}       
\usepackage{url}            
\usepackage{booktabs}       
\usepackage{amsfonts}       
\usepackage{nicefrac}       
\usepackage{microtype}      
\usepackage[table]{xcolor}
\usepackage{arydshln}

\usepackage{graphicx}
\usepackage{algorithm}
\usepackage{algpseudocode}
\usepackage{multirow}
\usepackage{makecell}
\usepackage{amsmath}

\usepackage{longtable}
\usepackage{array}
\usepackage{tabularx}

\usepackage{amssymb}
\usepackage{pifont}
\usepackage{tikz}
\usepackage{tcolorbox}
\usepackage{setspace}

\usepackage{graphicx,wrapfig}
\usepackage{enumitem,kantlipsum}
\hypersetup{colorlinks, citecolor=blue}

\title{Finding "Good Views" of Electrocardiogram Signals for Inferring Abnormalities in Cardiac Condition}

\author{%
  Hyewon Jeong \\
  MIT EECS, CSAIL\\
  \texttt{hyewonj@mit.edu}\\
   \And
   Suyeol Yun \\
   MIT Political Science \\
   \texttt{syyun@mit.edu} \\
   \And
   Hammaad Adam \\
   MIT IDSS \\
   \texttt{hadam@mit.edu} \\
}

\usepackage{etoolbox}

\BeforeBeginEnvironment{figure}{\vskip-1.4ex}
\AfterEndEnvironment{figure}{\vskip-1.4ex}

\begin{document}

\maketitle
\begin{abstract}
Electrocardiograms (ECGs) are an established technique to screen for abnormal cardiac signals. Recent work has established that it is possible to detect arrhythmia directly from the ECG signal using deep learning algorithms. While a few prior approaches with contrastive learning have been successful, the best way to define a positive sample remains an open question. In this project, we investigate several ways to define positive samples, and assess which approach yields the best performance in a downstream task of classifying arrhythmia. We explore spatiotemporal invariances, generic augmentations, demographic similarities, cardiac rhythms, and wave attributes of ECG as potential ways to match positive samples. We then evaluate each strategy with downstream task performance, and find that learned representations invariant to patient identity are powerful in arrhythmia detection. We made our code available in \url{https://github.com/mandiehyewon/goodviews_ecg.git}
\end{abstract}

\section{Introduction}

Electrocardiogram (ECG) has been used as a routine diagnostic method to screen out abnormal cardiac signals and underlying heart conditions of a patient for decades. As an ECG signal inherently contains the hemodynamics of a patient, there have been various attempts to diagnose the patient's condition with an ECG signal \cite{hannun2019cardiologist, attia2019screening, attia2019artificial, tison2019automated}. However, many current machine learning methods require costly, high quality labels, and fail to utilize vast unlabeled datasets.

Recent works on self-supervised contrastive learning address this problem by leveraging unlabeled datasets to build useful representations for mainstream tasks \cite{chen2020simple, he2020momentum}. Contrastive learning thus offers a compelling way to learn ECG representations that yield accurate downstream predictions \cite{diamant2021patient, gopal20213kg, kiyasseh2021clocs}. While a few prior studies have adopted this approach, the best way to define a positive sample for contrastive learning in this setting is still an open question \cite{tian2020makes}. Given an ECG, should a corresponding positive sample be an augmented version \cite{gopal20213kg} or projection \cite{kiyasseh2021clocs} of the same ECG test? Is it better to use a different test from the same patient? \cite{diamant2021patient} Or perhaps even a scan from a different patient with a similar demographic background and clinical history? Could the distance between the ECG signals dictate the similarity of the patient's heart function \cite{jeong2023deep}?

Optimal views for contrastive learning are the ones that retain task-relevant information, while retaining information invariant to views \cite{arjovsky2019invariant} and minimizing the mutual information between views \cite{tian2020makes}. For a given ECG, previous works have constructed positive samples either by transforming the same scan \cite{kiyasseh2021clocs} or by using another scan from the same patient \cite{diamant2021patient}. However, the first approach is scan-specific and may contain less than ideal variance, while the second requires multiple scans per patient, which may not always be available. Therefore, we investigate whether it is possible to leverage ECGs from different patients when defining a view for contrastive learning. Such an approach would produce representations that are invariant to the specific identity of the patient; as clinical diagnoses are shared across broader groups of patients, patient-invariant representations may prove to be more accurate for downstream clinical diagnosis.

There are several previous works showing that ECG signals have the demographic information of a patient, where the models could classify the racial \cite{noseworthy2020assessing} and gender information \cite{attia2019age} out of ECG signals. Also, wave information automatically extractable from the ECG signals (PR interval, QRS interval, QT interval, etc.) can also help a downstream task detecting different types of abnormalities. We leverage these demographic and wave similarities as well to define a set of strategies for producing patient-invariant representations. We evaluate these representations on a downstream task of detecting different types of arrhythmias, and compare their performance to patient- and signal-specific representations from prior works \cite{kiyasseh2021clocs, chen2020simple}.

Our work has three key contributions. First, as established in prior research \cite{kiyasseh2021clocs,diamant2021patient}, we find that contrastive learning is an effective way of learning ECG representations; all methods we consider outperform a supervised baseline. Second, we determine that patient-invariant representations are more accurate than existing patient-specific ones, demonstrating the power of pooling information across similar patients. Finally, we propose a novel matching strategy based on automatically extracted ECG wave attributes. This approach is simple and effective and proves the most effective way to learn representations from unlabelled ECG data.

\begin{figure*}[t!]
  \centering
  \includegraphics[width=1.0\textwidth]{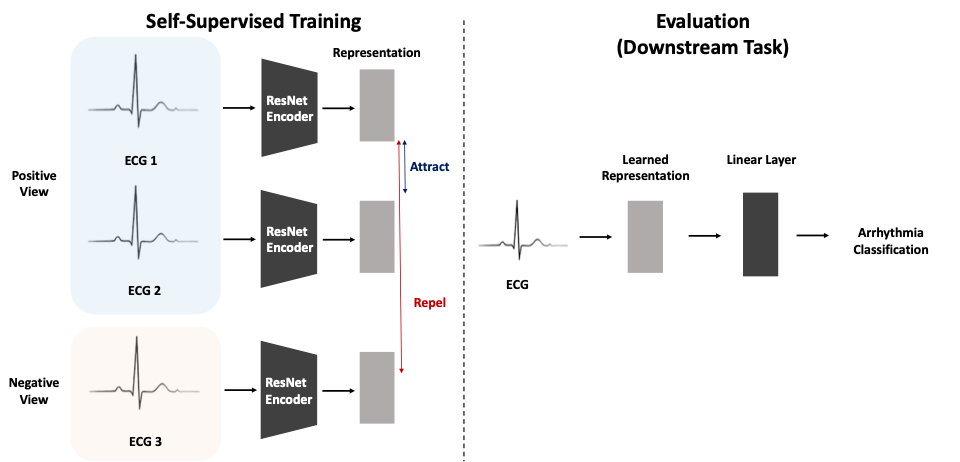}
  \caption{\textbf{Concept}. ECG 1, ECG 2, and ECG 3 on the left are sampled according to the strategies in Section \ref{sec2.1} (e.g., given ECG 1 drawn from the male subgroup (strategy 3), we can sample ECG 2 from the same gender group and ECG 3 from distinct gender group). The contrastive learning objective attracts the learned representation from positive samples and repels those from negative samples. We then evaluate the learned representation on the downstream task.} 
\label{concept}
\end{figure*}

\section{Approach}
Our approach has two key steps. First, we use a number of different strategies -- both patient-specific and patient-invariant -- to define a positive sample, and use each of these to contrastively learn a representation of an ECG. We then evaluate the quality of these representations using a downstream classification task. The specific task we chose in our project is to classify three major classes of arrhythmias with one normal cardiac condition class (Table \ref{arr_chapman}). This task is clinically impactful as we can triage patients to identify four major classes of cardiac conditions using the unseen ECG of new patient.

\subsection{Self-Supervised Contrastive Learning}\label{sec2.1}
We first provide a quick formulation of the contrastive learning problem. Let $D = \{(x_i, y_i)| i \in [N] \}$ be a dataset with medical diagnosis label $y_i$ for each ECG signal $x_i$. Our goal is to learn an encoder function $f$ such that an ECG's latent representation  $z_i = f(x_i)$ captures its key clinical information and can be used for downstream prediction. To do so, for each scan $x_i$, we define a set of views $S^{\texttt{+}}_{\tilde{x}_i} = \{ \tilde{x}_j | \text{ } (x_i, \tilde{x}_j) \text{ is a positive pair} \}$ according to some predetermined strategy. These $\tilde{x}_j$ can either be transformed versions of $x_i$ itself or other scans from $D$ (with or without transformation). Similarly, we define $S^{\texttt{-}}_{{x}_i}= \{ \tilde{x}_j | \text{ } (x_i, \tilde{x}_j) \text{ is a negative pair} \}$ and $S_{{x}_i} =S^{\texttt{+}}_{{x}_i} \cup S^{\texttt{-}}_{{x}_i}$. Intuitively, $S^{\texttt{+}}_{{x}_i}$ should include examples that are clinically similar to $x_i$, while $S^{\texttt{-}}_{{x}_i}$ should include dissimilar examples. Then, the contrastive learning approach learns the optimal representations $z^*_{{x}_i} = f^*({x}_i)$ by minimizing the following loss function
\begin{align*}
\mathcal{L} = \sum_{i}^N-\log \frac{\sum_{S^{\texttt{+}}_{\tilde{x}_i}} \exp \left[\operatorname{sim}\left(f(x_i), f(\tilde{x}_j)\right) / \tau\right]}{\sum_{S_{\tilde{x}_i}}\exp \left[\operatorname{sim}\left(f(x_i), f(\tilde{x}_j\right) / \tau\right]}
\end{align*}
where $\operatorname{sim}$ denotes cosine similarity, $\tau$ denotes a temperature parameter, and $f$ is parametrized using a Resnet Encoder (Fig. \ref{concept}).

\subsection{Defining Transformation-Invariant Positive Sample}
\label{sec:defpos}
Effectively, our choice of $S^{\texttt{+}}_{\tilde{x}_i}$ define what transformations the learned representations $z$ should be invariant to. Consider a simple case where $\mathcal{T}$ is a transformation that adds Gaussian noise to $x_i$ and $S^{\texttt{+}}_{{x_i}} = \{\mathcal{T}(x_i)\}$. The contrastive approach thus requires the optimal $f$ be (approximately) invariant to $\mathcal{T}$, that is, $f(x_i) \approx f(\mathcal{T}(x_i))$. This strategy posits that Gaussian noise shouldn't change the clinical information conveyed by the scan, and thus adding noise should not affect the learned representations. Similarly, consider a different strategy; if $G_1...G_m$ is some partition of $D$ and $x_i \in G_k$, define $S^{\texttt{+}}_{{x_i}} = \{x_j \in G_k\}$ and $S^{\texttt{-}}_{{x_i}} = \{x_j \in G_k^c\}$. The contrastive approach thus posits that if $x_i, x_j \in G_k$, then $f(x_i) \approx f(x_j)$. This fact is what we refer to as patient invariance: the encoded representation should depend on a patient's group membership, but not on their specific identity. 

Given this formulation, we propose 5 strategies regarding how to define $S_{{x}_i}$. Two of these (Strategy 1 and Strategy 2) learn patient-specific representations, while the other three can learn patient-invariant representations. The patient-specific strategies assume that relevant medical context is shared at a patient level, while the patient-invariant strategies assume that this context is shared more broadly within groups of patients. In our experiments, we will evaluate the representations produced by each of these approaches on a downstream arrhythmia classification task, and compare results to see whether it is actually possible to learn meaningful patient-invariant representations. 

\textbf{Strategy 1: Using temporal and spatial invariance of the same ECG (e.g. CLOCs \cite{kiyasseh2021clocs})}\\
As ECG is a \textit{repeating} signal recorded from a different \textit{location} of the heart, we can exploit both temporal and spatial information present in the ECG. So our strategy is to capture the high-level context of medical diagnosis where representations become invariant to temporal and spatial differences of the same ECG. Specifically, context changes of the ECG are unlikely to occur in short duration, so we can pair adjacent segments of short duration as positive. Formally speaking, we divide $x_i$ with duration of $S$ ($5,000$ timestamps in Chapman dataset) into $V$ number of non-overlapping temporal segments. We assume that context doesn't change during this temporal window with the length of $S/V$ and define $T_{time}$ which samples any non-overlapping pairs in the same temporal window. Then we sampled two different $T_{time}(x_i)$s and put them into $S^{\texttt{+}}_{\tilde{x}_i}$. In addition, recordings from different leads at the same time share the spatially invariant context, thus those can be paired as positive samples as well. We can define $T_{lead}$ which samples any one of 12 leads from $x_i$ and can include any pair of $T_{lead}(x_i)$s into $S^{\texttt{+}}_{\tilde{x}_i}$.

\textbf{Strategy 2: Using generic transformations of the same ECG  (e.g. SimCLR \cite{chen2020simple})}\\
A pair of the same ECG signal can be naturally understood as sharing the same context. Therefore we can try to learn representations of ECGs by  maximizing agreement between differently augmented views of the same data example. Specifically, we perform a stochastic data augmentation where two different augmentations $T_1, T_2$ are randomly selected from $\mathcal{T}$ and apply $T_1, T_2$ to $x_i$. This will result in two correlated views of the $x_i$, which we consider as a positive pair. So we include $(T_1(x_i), T_2(x_i))$ into $S^{\texttt{+}}_{\tilde{x}_i}$. For this strategy, we include generic augmentation techniques for time series data, such as jittering, scaling, flipping, rotating, time/size-warping \cite{park2019specaugment} in $\mathcal{T}$.

\textbf{Strategy 3: Matching on demographics}\\
A morbidity rate, which represents rates of acute and chronic diseases in a population, is usually reported for different demographics. This implies that demographics (e,g, age, gender, and race) can be understood as sharing the same high-level context of medical diagnosis. Therefore, we can randomly match two different ECG signals which share the same demographic features and consider them as a positive pair. Chapman dataset contains age and gender as demographic features for each $x_i$. We sample $x_j$ which falls into the same age and gender group with $x_i$ and include the pair $(x_i, x_j)$ into $S^{\texttt{+}}_{\tilde{x}_i}$.

\textbf{Strategy 4: Matching on cardiac rhythm}\\
Cardiac rhythms interpreted from the ECG can be classified as normal, slow, irregular, etc by medical experts. In addition, medical experts can classify the condition of the heart as blocked, axis of contraction being shifted, tissue being enlarged, etc via interpretation of ECG. Therefore, with a discrete label of cardiac rhythms and conditions inferred from the ECG by experts, we can define a positive pair of ECG which shares the same cardiac rhythms and conditions. Chapman dataset includes 11 and 16 discrete categories of cardiac rhythms and conditions for each $x_i$. We sample $x_j$ which falls into the same categories of cardiac rhythms and conditions with $x_i$ and include the pair $(x_i, x_j)$ into $S^{\texttt{+}}_{\tilde{x}_i}$.

\textbf{Strategy 5: Matching on wave attributes of ECG}\\
For each ECG signal, Chapman dataset includes 11 diagnostically meaningful attributes of ECG, such as ventricular rate, atrial rate, and QRS duration, etc. Since medical experts refer to these attributes for medical diagnosis, we can understand a pair ECG signals with similar attributes will share similar context in terms of diagnosis. 11 attributes are all represented as numeric so we can construct positive pairs by measuring the Euclidean distance between attributes considered as vectors. Let's denote $a_{x_i} \in \mathbb{R}^{11}$ which defines 11 attributes of $x_i$ as a vector. Then we sample $x_j$ which satisfies $\Vert a_{x_i} - a_{x_j}\Vert_2 \le h$ and include the pair $(x_i, x_j)$ into $S^{\texttt{+}}_{\tilde{x}_i}$. $h$ is a cutoff threshold for sampling.

\subsection{Evaluating learned representation on arrhythmia classification}

Our goal is to learn representations of ECGs that capture important information about a patient's cardiac health. We thus assess the utility of the learned representations using accuracy on a downstream arrhythmia classification task (Figure \ref{concept}). Each ECG falls into one of four rhythm categories: sinus rhythm (normal), sinus bradycardia (slow heartbeat), supraventricular tachycardia (SVT, fast heartbeat in supraventricular area), and atrial fibrillation/flutter (irregular or rapid heartbeat). For each of the contrastive strategies discussed above, we train a linear classifier to predict a patient's cardiac rhythm from the learned representations. We then evaluate the strategy using the predictive accuracy of this linear model on held-out test data. The higher the accuracy, the better the strategy, as the  more information its representations capture about a patient's cardiac health. As a baseline, we also test the performance of a CNN classifier that predicts rhythm directly from an ECG (i.e. no representation). Note that this choice of evaluation task is consistent with prior work \cite{kiyasseh2021clocs}. 

\section{Experiments}

\subsection{Dataset}
\label{sec:dataset}
We experiment on the publicly available Chapman dataset, which includes 12-lead ECG \cite{zheng2020optimal, zheng202012} of 10,369 patients for arrhythmia classification. The ECG signal data was sampled with 500 Hz of sampling rate and the whole sample size is 10 seconds of ECG streak (total 12-dimensional channel input with 5000 time points). The data includes 5,804 male patients and 4,565 female patients and includes 542 / 702 / 1,254 / 2,041 / 2,614 / 1,863 / 1,353 patients for each age group of 18-29 / 30s / 40s / 50s / 60s / 70s / $>$80s. We merge the $11$ rhythm information into four groups (Table \ref{arr_chapman}) as suggested in \cite{zheng202012}. These groups correspond to 3 major arrhythmia classes and 1 normal condition in order. We randomly split the dataset into train, test and valid with a ratio of 60\%/20\%/20\%.

\subsection{Experimental Results}
Table \ref{Result} summarizes the experimental results of the five contrastive strategies and supervised baseline. We focus our discussion on four key takeaways; further details on experimental setup and implementation are provided in Appendix \ref{appendix:expdetails}.

\begin{table*}[ht!]
	\small
	\centering
	\begin{tabular}{c|cc}
		\toprule
		Methods & AUROC \\
		\midrule
		Supervised Baseline & 81.4 $\pm$ 1.3 &\\
		\midrule
		Strategy 1 (CLOCS \cite{kiyasseh2021clocs} - Lead) & 88.9 $\pm$ 1.6 &\\
		Strategy 1 (CLOCS \cite{kiyasseh2021clocs} - Time) & 83.0 $\pm$ 1.1 &\\
		Strategy 2 (SimCLR \cite{chen2020simple})&  89.6 $\pm$ 0.6 & \\
		Strategy 3 (Demographics) & 88.1 $\pm$ 1.3 &  \\
		\textbf{Strategy 4 (Rhythms)} & \textbf{99.3 $\pm$ 0.0} &   \\
		\midrule
		\textbf{Strategy 5 (Attributes)} & \textbf{94.2 $\pm$ 0.3} &   \\
		\bottomrule
	\end{tabular}
\caption{\textbf{Result} of each strategy trained for $50$ epochs with metric AUROC. Results were averaged over 5 runs.}
\label{Result}
\end{table*}

\textbf{Self-supervised learning is extremely effective.} The result in Table \ref{Result} shows that Supervised baseline
scores AUROC = $84.1$ which is lower than all other strategies. The result implies that using contrastive learning does contribute to higher performance of downstream tasks as shown in prior works  \cite{diamant2021patient, gopal20213kg, kiyasseh2021clocs}. In addition, Strategy 1 (CLOCS), Strategy 2 (SimCLR), Strategy 3 (Demographics), and Strategy 5 (Attributes) are trained without any particular labels on ECG. Therefore, this result extends the effectiveness of contrastive learning to unlabeled self-supervised settings, especially in the clinical context of routine screening tests for arrhythmia detection.

\textbf{Patient-invariant representations are as or more powerful than patient-specific ones}. The result in Table \ref{Result} shows that Strategy 1 (CLOCS) scored AUROC = 85.95 on average for Time and Lead and Strategy 2 (SimCLR) scored AUROC = 89.6. These two strategies are slightly better than Strategy 3 (Demographics) and lower than Strategy 4 (Rhythms) and 5 (Attributes) which scored AUROC 99.3 and 94.2, respectively. The result shows that Strategies 3, 4, and 5 show better performance than Strategies 1 and 2 on average. As Strategy 1 (CLOCS) and 2 (SimCLR) fall in patient-specific self-supervised learning while other strategies fall in patient-invariant learning as explained in section \ref{sec:defpos}, our result shows that patient-invariant learning can be an effective way of extracting diagnostic information which is shared among the same group of patients. 

\textbf{Matching on the downstream label produces extremely accurate results but is of limited use.} The result in Table \ref{Result} shows that Strategy 5 (Rhythms) scores best performance of AUROC 99.3,  higher than all other strategies. For the interpretation of this result, we hypothetically conclude that if the matching of positive-negative samples is more closely related to the downstream task, it tends to record higher performance. This is because classifying different types of arrhythmia is  performed by directly observing
(Cardiac) Rhythms by medical practitioners in reality as we defined in Strategy 4. However, this approach is of limited use: it requires labeled data and doesn't generalize to the semi-supervised setting. Moreover, it is possible that such representations may be overfit to our specific choice of downstream task: testing this is an important step for future work.

\textbf{Matching on wave attributes is both highly effective and useful.} The result in Table \ref{Result} shows that Strategy 5 (Attributes) scores the second-best performance of AUROC 94.2 following the best performance of Strategy 4 with AUROC = 99.3. Although Strategy 5 (Attributes) shows less performance than Strategy 4 (Rhythms), since wave attributes can be automatically computed and provide useful information to numerically describe the cardiac conditions, it's more practical than Strategy 4 (Rhythms) which requires semantic labels from medical experts. As the result of Strategy 5 (Attributes) scores comparably higher than any other strategies other than Strategy 4 and slightly lower than Strategy 4, we can conclude that wave attribute provides a "good view" of matching a positive sample to produce effective representations to detect arrhythmia without any dependence on costly labels from medical experts. 

\section{Conclusion}
Our experiment and result provides three important findings.  First, self-supervised learning without any labels from medical experts can contribute significantly to producing effective representations to detect arrhythmia. Second, wave attributes of ECG as a view of matching a positive sample is especially useful in terms of its relatively higher performance than other strategies on arrhythmia classification task and zero-dependency on labels from medical experts. Third, views of matching a positive sample tends to be more effective if the view can provide patient-invariant information which generalizes clinically important features shared across patients with the same disease.

Given more time, we would carry out a more extensive testing of the five strategies. Specifically, it is important to determine whether our results generalize to different datasets and different downstream classification tasks. We would also like to evaluate the representations beyond their use in linear classification. For example, can the learned representations be useful in transfer learning? Finally, while this paper suggests which methods work best, it does not establish why. Interpreting the representations produced, through both automated methods like GradCAM \cite{selvaraju2017grad} or feature-level attention \cite{vaswani2017attention}  and actual clinical evaluation with medical practitioners, is an important direction for future work.

\section{Acknowledgements and Author Contributions}
This is a research project done as a part of the Machine Learning 6.867 class 2021 Fall at MIT. We thank Yumeng Cao and Professor Tommi Jaakkola for their insightful comments and discussion.

All three students equally devoted a significant amount of time and effort to this project.

\textit{Hyewon Jeong} came up with the idea of finding invariant views for downstream arrhythmia classification. She interpreted the dataset and implemented a dataloder for the Chapman dataset. She also implemented the whole pretext and downstream training loop, main skeleton of code, Supervised Baseline, and Strategy 2 (SimCLR), and ran to fine-tune all Baselines: Supervised baseline, Strategy 1 through 5.

\textit{Suyeol Yun} formulated mathematical notations in the approach section that can encompass all different strategies. He implemented Contrastive Loss, Strategy 2 (SimCLR), and Strategy 5 (Wave Attributes). He also implemented and debugged the whole pretext training loop and some parts of the downstream training loop. 

\textit{Hammaad Adam} formalized the concept of patient-invariance and patient-specific identity and Strategy 5. He designed a way of computing contrastive loss with a shared code block across different strategies. He interpreted the dataset and implemented a dataloader for the Chapman dataset. He also implemented Strategy 1 (CLOCS), Strategy 3 (Demographic), Strategy 4 (Cardiac Rhythm), and the training loop for the downstream classification task.

\bibliographystyle{plain}
\bibliography{reference}

\medskip

\newpage
\section{Experimental Details}
\label{appendix:expdetails}

\subsection{Major Arrhythmia Classes}
\label{arr_class}
We summarize the number of instances and information of each downsteram class rhythms in Table \ref{arr_chapman}. We merge several rhythm condition to total four classes. 

\begin{table*}[ht!]
	\small
	\centering
	\begin{tabular}{c|cc}
		\toprule
		Class Name & Conditions & Total size \\
		\midrule
		AFIB & AFIB, AF  & 2,218\\
		GSVT & SVT, AT, SAAWR, ST,
AVNRT, AVRT & 2,199\\
		SB & SB & 3,871\\
		SR & SR & 2,081 \\
		\midrule
		All & All & 10,369  \\
		\bottomrule
	\end{tabular}
\caption{Arrhythmia classifications used for downstream task, adopted from \cite{zheng202012}}
\label{arr_chapman}
\end{table*}

\subsection{Pre-training Implementation}
\label{sec:pretrain}
We conducted a pre-training experiment on the training dataset explained in \ref{sec:dataset}. We have used ResNet-18, a 18 layer ResNet \cite{resnet} 
 architecture as our  Resnet Encoder in Figure \ref{concept}. Resent Encoder $f(\cdot)$, produces a representation of each 12-lead ECG. Given an ECG $x_i$, it outputs the encoding $f(x_i) = z_i$. We trained our Resnet Encoder to minimize the contrastive loss in \ref{sec2.1}.
\vspace{-2mm}
\subsection{Evaluation on Downstream task}
\label{sec:eval}
We evaluate our pre-trained representations with a downstream task of supervised classification. Each ECG $x_i$ is labeled with one of the 4 different arrhythmia classes as explained in \ref{sec:dataset}.
We used one layer of linear classifier $\mathbf{H} \in \mathbb{R}^{\text{embed dim} \times 4}$ to produces score distribution of label $\mathbf{H}(z_i)$. After we train $\mathbf{H}$ using cross entropy loss, we evaluated AUROC by comparing $\mathbf{H}(z_i)$ with the label of $x_i$.
\vspace{-2mm}
\subsection{Baseline}
To check the performance improvement through the pre-training, we implemented our baseline model using the same Resnet Encoder and linear classifier as explained in
\ref{sec:pretrain} and \ref{sec:eval}
\label{sec:baseline}. For the baseline model, we skip the pre-training step that minimizes contrastive loss but train the same Resnet Encoder and linear classifier only using cross-entropy loss and evaluate AUROC.
\vspace{-2mm}
\subsection{Hyperparameters}
\label{sec:hyperparams}
For pre-training, we chose the embedding dimension of $128$ and batch size of $64$. For the contrastive loss, we chose temperature parameter $\gamma = 0.07$. We trained our ResNet Encoder for $50$ epochs with a learning rate of $0.0001$ in the pre-training step in \ref{sec:pretrain} while we trained our linear classifier for $10$ epochs with a learning rate of $0.01$ for evaluation on the downstream task in \ref{sec:eval}. In addition, we trained the baseline model in \ref{sec:baseline} for 50 epochs with a learning rate of $0.0001$. For each step of pre-training, baseline, and evaluation, we used the batch size of $64$.

\end{document}